\documentclass[aps,twocolumn,showpacs,preprintnumbers,pra,superscriptaddress]{revtex4}
\usepackage{graphicx}
\usepackage{amssymb}
\usepackage{amsmath}

\newcommand{\R}{\mathbf{r}}
\newcommand{\Tr}[1]{\mathrm{Tr}\left[#1\right]}
\newenvironment{acknowledgment}{{\flushleft \bf Acknowledgments:}}{}

\begin{document}

\title{Relevance of coordinate and particle-number scaling in density functional theory}

\author{Eduardo Fabiano}
\affiliation{National Nanotechnology Laboratory (NNL), Istituto di Nanoscienze-CNR,
Via per Arnesano 16, I-73100 Lecce, Italy }
\author{Lucian A. Constantin}
\affiliation{Center for Biomolecular Nanotechnologies @UNILE, Istituto Italiano di Tecnologia, 
Via Barsanti, I-73010 Arnesano, Italy}

\date{\today}

\begin{abstract}
We discuss a $\beta$-dependent family of electronic density scalings 
of the form  $n_\lambda(\R)=\lambda^{3\beta+1}\; n(\lambda^\beta \R)$ in the 
context of density functional theory. In particular, we consider 
the following special cases: the Thomas-Fermi scaling 
($\beta=1/3$ and $\lambda \gg 1$), which is crucial for the semiclassical
theory of neutral atoms; the uniform-electron-gas scaling 
($\beta=-1/3$ and $\lambda\gg 1$), that is important in the semiclassical 
theory of metallic clusters; the homogeneous density scaling ($\beta=0$) 
which can be related to the self-interaction problem in density functional 
theory when $\lambda \leq 1$; the fractional scaling 
($\beta=1$ and $\lambda\leq 1$), that is important for atom and molecule
fragmentation; and the strong-correlation scaling ($\beta=-1$ and $\lambda \gg 1$) that
is important to describe the strong correlation limit.

The results of our work provide evidence for the importance of this 
family of scalings in semiclassical and quantum theory of electronic
systems, and indicate that these scaling properties must be considered as
important constraints in the construction of new approximate density
functionals. We also show, using the uniform-electron-gas scaling, 
that the curvature energy of metallic clusters is related to 
the second-order gradient expansion of kinetic and exchange-correlation 
energies.

\end{abstract}


\maketitle


\section{Introduction}
\label{sec1}
In Density Functional Theory (DFT) \cite{KS,bookdft,book2}, 
the density scaling is 
a key concept and was used along the years to derive many exact constraints 
\cite{PL1,LG1,zhao93,ivanov98,zhang98,chan99,toulouse05,nagy05,nagy11}, 
as well as useful virial relations 
\cite{gosh85,PL1,vanleeuven95,cruz98,levy09,gaiduk09,elkind12}
for the kinetic, exchange and correlation energy functionals. 
Furthermore, the density scaling has a fundamental role in DFT 
due to its intimate relation with the adiabatic connection formalism
\cite{PL1,levy91,wang97}.

One important family of scaling transformations for the particle density
$n$ is defined by the general linear transformation \cite{LG1}
\begin{equation}\label{e1}
n_{\mathit{M},\mathbf{a}}(\R) = \mathrm{det}(\mathit{M})\ n(\mathit{M}\R +\mathbf{a})\ ,
\end{equation}
where $\mathit{M}$ is a real invertible $3\times 3$ matrix and 
$\mathbf{a}\in\mathbb{R}^3$. Scaling transformations of this kind
correspond to changing the external potential associated with the
density whilst preserving the normalization of the density, i.e.
\begin{equation}\label{e2}
\int n_{\mathit{M},\mathbf{a}}(\R) d\R = \int n(\R) d\R = N\ .
\end{equation}
The most important scalings connected with Eq. (\ref{e1}) 
are the ones defined by 
\begin{equation}\label{e3}
\mathit{M}_{ij} = \lambda_j\ \delta_{ij}\; , \; \lambda_j> 0 \quad \mathrm{and}\quad \mathbf{a}=0\ ,
\end{equation}
i.e., with $\mathit{M}_{ij}$ a diagonal matrix with positive elements.
In particular, there are three cases of high physical interest:

(i) the \emph{uniform scaling} defined by 
$\lambda_1=\lambda_2=\lambda_3=\lambda \geq 0$, 
hence $n_\lambda(\R)=\lambda^3n(\lambda\R)$. 
Under uniform scaling the Kohn-Sham exchange and the
non-interacting kinetic energies transform as
$E_x[n_\lambda]=\lambda E_x[n]$ and $T_s[n_\lambda]=\lambda^2 T_s[n]$,
respectively \cite{PL1}. However, the interacting kinetic energy $T[n]$ and the
non-local Hartree-Fock exchange energy $E_x^{HF}[n]$ do not have these
properties \cite{PL1}. All popular exchange and kinetic energy functionals
are nowadays designed to satisfy the uniform scaling relation \cite{bookdft,book2}.

(ii) The \emph{two-dimensional nonuniform scaling of the density} defined 
by $\lambda_1=1$ and $\lambda_2=\lambda_3=\lambda \geq 0$, 
hence $n^{zy}_{\lambda}(x,y,z)=\lambda^2 n(x,\lambda y,\lambda z)$, 
under which the system approaches the 1D limit when $\lambda\rightarrow \infty$.
Although there are important conditions for both Kohn-Sham exchange and 
correlation energies under this scaling \cite{LG1}, 
they are not satisfied by any popular XC functional (to our knowledge).

(iii) The \emph{one-dimensional nonuniform scaling of the density} defined 
by $\lambda_1=\lambda_2=1$ and $\lambda_3=\lambda \geq 0$, 
hence $n^{z}_{\lambda}(x,y,z)=\lambda n(x,y,\lambda z)$, 
under which the system approaches the 2D limit when $\lambda\rightarrow \infty$.
This density scaling, that is related to the dimensional crossover
of the XC energy (from 3D to 2D) \cite{PoP,CPP,CLA}, has been recently
incorporated in a semilocal XC functional, 
named q2D-generalized-gradient-approximation (q2D-GGA) \cite{q2D}, 
constructed for
mild and strong quasi-2D regimes. The remarkable performance of the q2D-GGA 
for surface energies and lattice constants of transition metals, 
showed the power of the one-dimensional nonuniform scaling.  

In this paper we consider a different type of scaling for the
density and focus the attention on the family of scaling relations
of the form
\begin{equation}\label{e4}
n_\lambda(\R) = \lambda^{3\beta+1}n(\lambda^\beta\R), \;\;\;\; \lambda > 0 ,
\end{equation}
where $\beta$ is a parameter.
These scaling transformations differ from those defined by Eq. (\ref{e1})
in the fact that they do not only change the external potential associated
with the density $n$, but also provide a change in the particle number
($N\rightarrow \lambda N$).
Well known members of the scaling family defined by Eq. (\ref{e4}) are
the Thomas-Fermi \cite{ELCB,Lieb2} and the homogeneous
density scaling \cite{chan99}, which are relevant for the semiclassical theory 
of the many-electron, non-relativistic, neutral atom \cite{ELCB}, and for 
the self-interaction error \cite{chan99}, respectively.
(Unless otherwise stated, atomic units are used throughout,
i.e., $e^2=\hbar=m_e=1$.)

This article is organized as follow: in Section \ref{sec2} we
shortly review the fractional-particle density functional theory, that 
is the correct framework for the scalings of Eq. (\ref{e4});
in Section \ref{sec3} we present scalings properties of useful density 
functionals; Section \ref{sec4} is devoted to the physical 
properties of selected density scalings (for $\beta$=1/3, -1/3, 0, 1, and -1);
and in Section \ref{sec5} we briefly analyze the performance of popular 
density functionals for the above scalings. Finally, in Section \ref{sec6}
we summarize our conclusions. 


\section{Theory and definitions}
\label{sec2}
Under the scaling transformations defined by Eq. (\ref{e4}) the
normalization of the particle density is modified, thus the
number of particles in the system is varied as $N\rightarrow\lambda N$,
with $\lambda N$ being in general a non-integer number.
The conventional picture of DFT, based on the Hohemberg-Kohn theorems
\cite{hk64} and/or the Levy constrained search \cite{levy79}
is thus inappropriate in this case.
In fact, a proper definition of densities with a non-integer number 
of particles is required in this case \cite{PPLB,yang00,ayers08}.
In this work we achieve this through the introduction of 
ensembles densities within a zero-temperature grand canonical 
ensemble theory \cite{PPLB}. 

The central quantity to consider is therefore the density-matrix operator
\cite{book2}
\begin{equation}\label{e5}
\hat{\Gamma} = \sum_M\sum_ip_{Mi}|\Psi_{Mi}\rangle\langle\Psi_{Mi}|\ ,
\end{equation}
where the sums are extended over all possible particle numbers $M$ and
over all the states of the $M$-particle Hamiltonian, $|\Psi_{Mi}\rangle$
is the $i$-th eigenstate of the $M$-particle Hamiltonian and $p_{Mi}$ is
the probability weight to find the system in the eigenstate
$|\Psi_{Mi}\rangle$ (hence, we have $0\leq p_{Mi}\leq 1$ and
$\sum_{Mi}p_{Mi}=1$).
The expectation value of any operator $\hat{A}$ is consequently
obtained as $\Tr{\hat{\Gamma}\hat{A}}$. In particular, for the particle 
density we have 
\begin{equation}\label{e6}
n(\R) = \Tr{\hat{\Gamma}\hat{n}(\R)}=\sum_kp_kn_k(\R)\ ,
\end{equation}
where we defined the super-index $k=Mi$, $n_k$ is the pure-state density 
of the $M$-particle $i$-th state, and we used  
\begin{equation}\label{e7}
\hat{n}(\R) = \hat{\psi}^\dagger(\R)\hat{\psi}(\R)\quad ,\quad 
\hat{\psi}(\R) = \sum_j\phi_j(\R)\hat{a}_j,
\end{equation}
with $\phi_i$ and $\hat{a}_i$ the single-particle orbital and 
annihilation operator of state $i$, respectively.
The total number of particles is, according to Eq. (\ref{e6}), 
\begin{equation}\label{e8}
N = \sum_kp_kN_k \quad , \quad N_k = \int n_k(\R)d\R\ .
\end{equation}
Similarly, for the one-particle density-matrix operator we have
$\hat{\gamma}_1(\R_1,\R_2) = \hat{\psi}^\dagger(\R_1)\hat{\psi}(\R_2)$
and 
\begin{equation}\label{e9}
\gamma_1(\R_1,\R_2)=\Tr{\hat{\Gamma}\hat{\gamma}_1(\R_1,\R_2)}=
\sum_kp_k\gamma_{1k}(\R_1,\R_2)\ .
\end{equation}

The ensemble Hohemberg-Kohn universal functional and the
non-interacting kinetic energy are defined as \cite{book2}
\begin{eqnarray}
\label{e10}
F[n] & = & \min_{\hat{\Gamma}\rightarrow n}\Tr{\hat{\Gamma}\left(\hat{T}+\hat{V}_{ee}\right)},\\
\label{e11}
T_s[n] & = & \min_{\hat{\Gamma}\rightarrow n}\Tr{\hat{\Gamma}\hat{T}},
\end{eqnarray}
where $\hat{V}_{ee}$ is the electron-electron repulsion operator and
the kinetic energy operator is defined by 
\begin{equation}\label{e12}
\hat{T} = \int\left[\nabla^2_{\R_2}\hat{\gamma}_1(\R_1,\R_2)\right]\delta(\R_1-\R_2)d\R_1d\R_2\ .
\end{equation}
The electron-electron repulsion operator can be further decomposed into
Coulomb, exchange, and correlation contributions to yield
\begin{eqnarray}
\label{e13}
J & = & \frac{1}{2}\int\frac{n(\R_1)n(\R_2)}{|\R_1-\R_2|}d\R_1d\R_2, \\
\label{e14}
E_x[n] & = & \Tr{\hat{\Gamma}_{min}\hat{E}_x}, \\
\label{e15}
E_c[n] & = & \Tr{\hat{\Gamma}_{min}\hat{E}_c}\ ,
\end{eqnarray}
where $\hat{\Gamma}_{min}$ is the density-matrix operator minimizing either
$\hat{T} +\hat{V}_{ee}$ or $\hat{T}$, according to Eqs. (\ref{e10}) and
(\ref{e11}), and with the exchange energy operator defined as \cite{book2}
\begin{eqnarray}
\label{e17}
\hat{E}_{x} & = & - \frac{1}{2}\int\frac{|\hat{\gamma}_1(\R_1,\R_2)|^2}{|\R_1-\R_2|}d\R_1d\R_2\ ,
\end{eqnarray}
while no explicit expression is known for $\hat{E}_c$.
Note that all the traces can be easily evaluated by use of Eqs. 
(\ref{e6}) and (\ref{e9}), together with the
resolution of identity $\hat{I}=\sum_{Mi}|\Psi_{Mi}\rangle\langle\Psi_{Mi}|$.

Within the theoretical framework sketched above the scaling relations
of Eq. (\ref{e4}) can be interpreted as an uniform scaling of the
pure-state densities
($n_{k}(\R)\rightarrow\lambda^{3\beta}n_k(\lambda^\beta\R)$ $\forall k$)
accompanied by a remodulation of the statistical weights such that
the particle number is changed to $\lambda N$. Note that this latter
is in general a complicated 
transformation because it must
accomplish the required particle-number variation
preserving the correct density and without violating
the normalization conditions for the statistical weights
($0\leq p_{k}\leq 1$ and $\sum_{k}p_{k}=1$). In particular, the simple 
transformation $p_k\rightarrow\lambda p_k$ in general
is not a suitable transformation,  as it brings a violation of 
the normalization conditions. Nevertheless, one special case is 
when we consider a system with one electron or 
less and $\lambda \leq 1$.
In this case in fact the density of fractional
charge $q$ can be written $n_q(\R) = qn_1(\R) + (1-q)n_0(\R)$, where
$n_1$ and $n_0$ are the densities for one particle and zero particles,
respectively (the latter is of course identically zero everywhere).
The scaling transformation yields then
$n_{q\lambda} = \lambda q\lambda^{3\beta}n_1(\lambda^\beta\R) + (1-\lambda
q)\lambda^{3\beta}n_0(\lambda^\beta\R)$. Thus, in this special case
we have indeed $p\rightarrow\lambda p$ (all information concerning the $n_0$ term
can be neglected). This important result will be employed in next section
to derive exact scaling relations for the non-interacting kinetic
energy and the Kohn-Sham exchange for systems with fractional occupation.

More insight into the properties of the transformation
governing the statistical weights can be achieved by considering 
(in analogy with the usual adiabatic connection procedure)
the M-particle Hamiltonian with a local potential which gives the 
right pure-state densities to recover Eq. (\ref{e4}).
In this way the trasformation
of the statistical weights, connected to scaling transformations of the 
type defined in Eq. (\ref{e4}), can be defined explicitly in 
two separate cases.
When, upon scaling, the variation in the particle number is smaller 
than one (i.e., $1 \leq \lambda \leq (N+1)/N$ or similarly  
$(N-1)/N \leq \lambda \leq 1$), the statistical
weights change with $\lambda$ as   
\begin{equation}\label{etr}
p_N: 1\rightarrow 1-N(\lambda-1) \quad ; \quad p_{N+1}: 0\rightarrow N(\lambda-1),
\end{equation}
and consequently, the exact total energy functional $E[n]$
varies linearly with $\lambda$ \cite{PPLB}
\begin{equation}
E[n_\lambda] = \left(1-N(\lambda-1)\right)E[n_N]+N(\lambda-1)E[n_{N+1}]\ ,
\end{equation}
where $n_N$ and $n_{N+1}$ denote $N$- and $(N+1)$-densities.
When the variation in the particle number is larger than one
(i.e., the number of particles changes from $N$ to $N+L+\omega$,
with $L$ an integer and $|\omega|<1$), the statistical weights change
as
\begin{equation}
p_N:1 \rightarrow 0 \quad ; \quad p_{N+L}: 0\rightarrow 1-\omega \quad ; \quad p_{N+L+1}: 
0\rightarrow 
\omega\ .
\end{equation}
Therefore, the transformation can be seen as a chain 
of transformations like the one in Eq. (\ref{etr}) 
concerning successively $p_N$, $p_{N+1}$, $\ldots$, $p_{N+L}$, $p_{N+L+1}$.
As a consequence, the exact total energy functional 
will be described as a succession of straight lines. 
Thus, the derivative discontinuity \cite{PPLB} plays an important 
role for the scaling transformations, especially when $N$ and $L$ are 
finite. However, in the limit
$L\rightarrow \infty$ the role of derivative discontinuity 
is diminished and one can always consider $\omega=0$. In this case the
behavior of the system is closely related to the semiclassical physics, 
as shown in the next sections. Because of the complexity of this density 
scaling, in this paper we mainly consider only two extreme cases of interest:
when $N=1$ and $\lambda\leq 1$, and the semiclassical limit 
$\lambda\rightarrow \infty$.


\section{General scaling properties}
\label{sec3}
Using the definition of Eq. (\ref{e4}) we can immediately find the
scaling properties of several quantities which depend explicitly
on the particle density. This is the case, for example, 
of the the Coulomb energy and any external potential 
(e.g., $E_{nuc}=\int nv_{nuc}d\R$ with $v_{nuc}\propto1/r$). Hence,
\begin{eqnarray}
\label{e23}
J[n_\lambda] & = & \lambda^{\beta+2}J[n], \\
\label{e25}
E_{nuc}[n_\lambda] & = & \lambda^{\beta+1}E_{nuc}[n].
\end{eqnarray}

Another interesting quantity is the local Seitz parameter 
$r_s=[3/(4\pi n)]^{1/3}$ which scales as
\begin{equation}\label{e26}
r_{s\lambda}(\R) = \lambda^{-\beta-\frac{1}{3}}r_s\left(\lambda^\beta\R\right)\ .
\end{equation}
Therefore, for $\beta>-1/3$ we have that $\lambda\rightarrow\infty$ 
implies $r_{s\lambda}\rightarrow 0$, while $\lambda\rightarrow 0$ 
implies $r_{s\lambda}\rightarrow \infty$. Thus, the conditions 
$\lambda\rightarrow\infty$ and $\lambda\ll 1$ correspond to 
scalings to the high- and low-density limits, respectively. 
The opposite is true for $\beta<-1/3$. 
For the special case $\beta=1/3$ instead the
local Seitz parameter is independent on $\lambda$ and the
density regime cannot be modified by a scaling transformation.

On the other hand, the usual density
parameters $s=|\nabla n|/2 k_F n$, $q=\nabla^2 n/\{4(3\pi^{2})^{2/3}n^{5/3}\}$,
$t=|\nabla n|/2 k_s  n$, and $v=|\nabla n|/2 k_v n$, 
with $k_F=(3\pi^2n)^{1/3}$ being the local
Fermi wave-vector \cite{LM,PBW}, $k_s=(4k_F/\pi)^{1/2}$ being the
Thomas-Fermi screening wave-vector \cite{LM,PBW}, 
and $k_v=2 (3/(4\pi^4))^{1/18}n^{1/9}$ being the wave vector 
suitable for bonding and valence regions \cite{zeta2},
scale according to
\begin{eqnarray}
\label{e27}
s_\lambda(\R) = \lambda^{-\frac{1}{3}}s\left(\lambda^\beta\R\right) \ & , & 
\ q_\lambda(\R) = \lambda^{-\frac{2}{3}}q\left(\lambda^\beta\R\right)\ ,\\ 
\label{e28}
t_\lambda(\R)  = \lambda^{\frac{\beta}{2}-\frac{1}{6}}t\left(\lambda^\beta\R\right)\ & , & 
\ v_\lambda(\R) = \lambda^{\frac{2\beta}{3}-\frac{1}{9}}v\left(\lambda^\beta\R\right)\ .
\end{eqnarray}
Thus, the reduced gradient and Laplacian for exchange and kinetic 
energies ($s$ and $q$) are independent on $\beta$ so that the
slowly-varying density limit ($s,q\rightarrow 0$) is reached 
whenever $\lambda\rightarrow\infty$, while for $\lambda\ll 1$
a rapidly-varying density regime is always set up.
On the contrary the density parameters $t$ and $v$, which
are relevant for the correlation, have a dependence on $\beta$.
Therefore, they can describe different density regimes depending on the
actual value of the parameter $\beta$.

Because for any value of $\beta$ the reduced gradient for exchange
and kinetic energy $s$ and the reduced Laplacian $q$ become
small in the limit $\lambda\rightarrow \infty$, it is also interesting to
investigate the scaling behavior of local density approximations (LDA)
and gradient expansions for the non-interacting kinetic energy 
and the exchange energy. These expressions will in fact become almost
exact in the limit $\lambda\rightarrow \infty$.
The required scaling relations are given by the formulas
\begin{eqnarray}
\label{e29}
E_x^{LDA}[n_\lambda] & = & \lambda^{\beta+4/3}E_x^{LDA}[n], \\
\label{e30}
E_x^{GE2}[n_\lambda] & = & \lambda^{\beta+2/3}E_x^{GE2}[n], \\
\label{e31}
T_s^{LDA}[n_\lambda] & = & \lambda^{2\beta+5/3}T_s^{LDA}[n], \\
\label{e32}
T_s^{GE2}[n_\lambda] & = & (1/9)T_s^W[n_\lambda] =(1/9) \lambda^{2\beta+1}T_s^W[n], \\
\label{e33}
T_s^{GE4}[n_\lambda] & = & \lambda^{2\beta+1/3}T_s^{GE4}[n]\ ,
\end{eqnarray}
where $E_x^{GE2}$ is the second-order gradient correction (GE2) term of the
exchange energy \cite{AK}, $T_s^{GE2}$ is the second-order gradient correction term
of the non-interacting kinetic energy \cite{bookdft}, $T_s^{W}$ is the von Weizs\"{a}cker 
kinetic energy functional \cite{bookdft}, and $T_s^{GE4}$ is the 
fourth-order kinetic energy gradient expansion term (GE4) \cite{Ho,BJC}.
Using these expressions, as well as Eq. (\ref{e23}), it is possible to
provide an useful accurate approximation for the universal functional
of Eq. (\ref{e10}) in the slowly-varying limit ($\lambda\rightarrow\infty$):
\begin{eqnarray}
\nonumber
F[n_\lambda] & \approx & \lambda^{\beta+2}J[n]+\lambda^{2\beta+5/3}T_s^{LDA}[n]+\\
\nonumber
&&+\lambda^{2\beta+1}T_s^{GE2}[n]+\lambda^{2\beta+1/3}T_s^{GE4}[n]+\\
\label{e34}
&&+\lambda^{\beta+4/3}E_x^{LDA}[n]+\lambda^{\beta+2/3}E_x^{GE2}[n]\ .
\end{eqnarray}
In this formula we neglected correlation contributions.
In fact, the leading term of LDA correlation energy in 
the high-density limit is \cite{PW1} $E_c^{LDA}[n]\propto \int d\R n  \ln(r_s)$,
and does not respect any simple scaling; whereas the leading term in the low
density limit is \cite{PW1} $E_c^{LDA}[n]\propto -\int d\R n r_s^{-1}$,
and scales as $E_c^{LDA}[n_\lambda] = \lambda^{\beta+4/3}E_c^{LDA}[n]$.
Moreover, the second-order correction to the correlation
energy \cite{HL} scales always as
$E_c^{GE2}[n_\lambda] = \lambda^{\beta+2/3}E_c^{GE2}[n]$.
However, due to the dependence on $\beta$ of the reduced gradients 
$t$ and $v$, the slowly-varying limit is only reached for $\beta<1/6$.
In addition, for any  $\beta>-1/3$ the scaling to the slowly-varying
limit corresponds also to a scaling to the high-density limit
($r_{s\lambda}\rightarrow 0$ when $\lambda\rightarrow\infty$).
In this limit the correlation contributions are negligible
with respect to the exchange part, and thus Eq. 
(\ref{e34}) becomes almost exact.
For $\beta\leq -1/3$ however correlation corrections to Eq. (\ref{e34})
might be needed.

To conclude this section we consider briefly the 
cases of Kohn-Sham kinetic and exchange energies. 
Using the convexity arguments derived in Ref. \cite{chan99},
and taking into account that the scaling family of Eq. (\ref{e4})
can be seen as an uniform scaling followed by a homogeneous scaling 
(i.e. $n_\lambda(\R)=\lambda (\lambda^{3\beta}n(\lambda^\beta\R))$), 
the following inequalities hold: 
\begin{eqnarray}\label{e22}
T_s[n_\lambda]\geq \lambda^{2\beta+1}T_s[n], \;\;\; \lambda>1,\\
T_s[n_\lambda]\leq \lambda^{2\beta+1}T_s[n], \;\;\; \lambda<1,
\end{eqnarray}
and
\begin{eqnarray}\label{e24}
|E_x[n_\lambda]|  \geq  \lambda^{\beta+1}|E_x[n]|, \;\;\; \lambda>1,\\
\label{eee22}
|E_x[n_\lambda]|  \leq  \lambda^{\beta+1}|E_x[n]|, \;\;\; \lambda<1.
\end{eqnarray}
Moreover, using the rigorous bound $T_s\geq T_s^W$, and the one conjectured 
by Lieb \cite{Lieb2} $T_s\leq T_s^{LDA}+T_s^W$ (for a rigorous, and tighter
upper bound of $T_s$ in terms of $T_s^{LDA}$, $T_s^W$ and $N$, see Eq. (23) of 
Ref. \cite{KEbound}), we can easily derive the following inequalities
\begin{equation}\label{ew22}
\lambda^{2\beta+1}T_s^W[n]\leq T_s[n_\lambda]\leq 
\lambda^{2\beta+5/3}T_s^{LDA}[n]+\lambda^{2\beta+1}T_s^W[n],
\end{equation}
for any $\lambda$ and $\beta$. In the case of exchange energy,
the Lieb-Oxford bound  \cite{LiebO1,LiebO2,LiebO3} 
$E_x\geq E_{xc}\geq 2.27 E_x^{LDA}$ gives
\begin{equation}\label{ew23}
E_x[n_\lambda]\geq 2.27 \lambda^{\beta+4/3}E_x^{LDA}[n].
\end{equation}

In the special case of one particle or less ($N\leq 1$)
and $\lambda\leq 1$,
using the formalism presented in Section \ref{sec2} (especially 
the fact that $p\rightarrow \lambda p$) it  can be shown
(see Appendix \ref{appa}) that
\begin{equation}\label{ee22}
T_s[n_\lambda]= \lambda^{2\beta+1}T_s[n], 
\end{equation}
and
\begin{equation}\label{ee24}
E_x[n_\lambda]= \lambda^{\beta+2} E_x[n]. 
\end{equation}
These results correctly agree with the scalings of $J[n]$ and $T_s^W[n]$, because 
in case $\lambda N\leq 1$, $E_x[n]=-J[n]$ and $T_s[n]= T_s^W[n]$ \cite{PS1}.
(Note also that Eq. (\ref{ee22}) holds for $N\leq 2$ \cite{PS1}.)  


\section{Selected scaling relations}
\label{sec4}
In this section we analyze in more detail the scaling relations corresponding
to special values of the parameter $\beta$. In this way we can highlight the
physical significance of the family of scaling transformations defined by Eq. 
(\ref{e4}) and provide evidence for its importance in electronic structure 
theory.


\subsection{Thomas-Fermi scaling ($\beta=1/3$)}
If we require the LDA non-interacting kinetic energy to scale as
the Coulomb energy, we find $\beta=1/3$, which corresponds to
the well known Thomas-Fermi scaling \cite{ELCB}.
With this choice the scaling of the density and the reduced gradients are
\begin{eqnarray}
\label{e35}
n_\lambda(\R) = \lambda^2n(\lambda^{1/3}\R) \ & , & \ r_{s\lambda}(\R) = \lambda^{-2/3}r_s(\lambda^{1/3}\R),\\
\label{e36}
s_\lambda(\R) = \lambda^{-1/3}s(\lambda^{1/3}\R) \ & , & \ q_\lambda(\R) = \lambda^{-2/3}q(\lambda^{1/3}\R)\ ,\\ 
\label{e37}
t_\lambda(\R)  = t(\lambda^{1/3}\R)\ & , & \ v_\lambda(\R) = \lambda^{1/9}v(\lambda^{1/3}\R)\ .
\end{eqnarray}
Thus, for $\lambda\rightarrow\infty$ the high-density limit is reached and
the exchange and kinetic energies, whose behavior is controlled 
by the density parameters $s$ and $q$, are in a slowly-varying density regime.
Therefore, the universal functional can be written \cite{ELCB}
\begin{eqnarray}
\nonumber
F[n_\lambda] &\approx& \lambda^{7/3}(T_s^{LDA}[n] + J[n])+\\
\nonumber
&&+\lambda^{5/3}(T_s^{GE2}[n]+E_x^{LDA}[n])+\\
\label{e38}
&&+\lambda(T_s^{GE4}[n]+E_x^{GE2}[n])\ ,
\end{eqnarray}
and the Thomas-Fermi kinetic energy ($T_s^{LDA}$) 
and the classical Coulomb energy
are the leading terms
in the total electronic energy ($E_{nuc}$ scales as $\lambda^{4/3}$).
This result is very important. In fact, the 
semiclassical theory of the many-electron neutral atom 
\cite{Sa80,ESa84,En88} is based on Eq. (\ref{e38}), which leads
to the semiclassical asymptotic expansion for the kinetic energy
\begin{equation}
T_s = c_0 N^{7/3}+c_1N^2+c_2N^{5/3}+\ldots\quad;\quad N\propto\lambda
\label{eee}
\end{equation}
and a similar one for exchange \cite{EB09}, 
that are very accurate (typical error of order 0.5\% - 0.2\%
even for real atoms \cite{ELCB,EB09,LCPB}). 
Note that the second term in the kinetic energy expansion ($c_1\lambda^2$),
can not be captured by the Thomas-Fermi scaling, being a quantum correction. 
Recently, it has been 
demonstrated that these asymptotic expansions are also important 
tools in DFT \cite{ELCB,APBE,APBEK,mukappa}, as they have been used to 
construct accurate non-empirical exchange-correlation \cite{APBE} and 
kinetic \cite{APBEK} energy functionals.

Extensions of Eq. (\ref{eee}) have been also proposed for
general ions and atoms \cite{march72,march80,tal82,ayers02},
which constitute a more challenging problem than neutral atoms alone. 
However, even for the first ionization potential of many-electron atoms, 
the extented semiclassical
Thomas-Fermi theory shows serious drawbacks and limitations \cite{ioni}, and 
accurate results can be obtained only within Kohn-Sham DFT or other 
orbital-dependent schemes.

Concerning correlation, we can obtain some insight by considering that,
any reasonable generalized gradient correction to the LDA correlation must be
designed to cancel the logarithmic divergence of the LDA term
under uniform scaling to the high density limit \cite{PBE}. 
Thus, in the high-density
limit $\epsilon_c^{GGA}\propto \ln\left(t^2\right)$. However, under the
Thomas-Fermi scaling , while $v_\lambda = \lambda^{1/9}v$, so 
that correctly $v\rightarrow\infty$ for
$\lambda\rightarrow\infty$ ($v$ is a density parameter suitable for 
valence and tail regions that are evanescent in a many-electron neutral atom),
the density parameter $t$ is just independent on $\lambda$. 
Hence, the gradient corrections to the correlation are independent on
the scaling and the whole correlation energy is dominated by the LDA
contribution $\epsilon_c^{LDA}\propto\ln(r_s)$ for $\lambda\rightarrow\infty$.
As a consequence, popular GGA functionals (e.g. PBE \cite{PBE}) that
recover LDA correlation in this limit, can be argued to be 
accurate (exact) \cite{ELCB}.

Finally, we mention the importance of the Thomas-Fermi scaling also 
for the atomic densities. In fact, under a Thomas-Fermi scaling to 
the high-density limit, the hydrogenic density $n=\exp(-2r)/\pi$
resembles features of the Thomas-Fermi density \cite{Lieb1}, becoming
slowly-varying over a Fermi wavelength (but not over the 
screening length $2\pi/k_s$). We recall that the Thomas-Fermi 
density, even if does not show shell structure and does not 
decay correctly, is a very good model for the densities of heavy atoms 
\cite{lieb77,lieb_arxiv,parr86,LCPB}. 
For an excellent discussion, see Ref. \onlinecite{ELCB}.


\subsection{Uniform-electron-gas scaling ($\beta=-1/3$)}
Consider a generic density parameter $d\propto|\nabla n|/n^\alpha$.
We define the uniform-electron-gas (UEG) scaling, as the scaling
belonging to the family of Eq. (\ref{e4}) that makes $d$ small for
any value of $\alpha$ in the limit $\lambda\rightarrow\infty$.
It is easy to prove that this scaling is defined by the parameter
$\beta=-1/3$.
Under such a scaling the density and the reduced gradients behave as
\begin{eqnarray}
\label{e39}
n_\lambda(\R) = n(\lambda^{-1/3}\R) \ & , & \ r_{s\lambda}(\R) = r_s(\lambda^{-1/3}\R)\\
\label{e40}
s_\lambda(\R) = \lambda^{-1/3}s(\lambda^{-1/3}\R) \ & , & \ q_\lambda(\R) = \lambda^{-2/3}q(\lambda^{-1/3}\R)\ ,\\ 
\label{e41}
t_\lambda(\R)  = \lambda^{-1/3}t(\lambda^{-1/3}\R)\ & , & \ v_\lambda(\R) = \lambda^{-1/3}v(\lambda^{-1/3}\R)\ .
\end{eqnarray}
Thus, both the density and the local Seitz parameter are independent
on $\lambda$ (except for a coordinate scaling), so that the scaling
does not involve any transformation towards the high- or low-density
limit. On the contrary, by construction, all the density parameters
vanish in the limit $\lambda\rightarrow\infty$, so the
slowly-varying density limit is fully recovered in this case
(hence, the name uniform-electron-gas scaling). Interestingly,
Eqs. (\ref{e40}) and (\ref{e41}) also show that for first-order
density parameters (i.e. those depending on $\nabla n$) exactly
the same dependence on $\lambda^{-1/3}$ is found under the 
uniform-electron-gas scaling.

For large values of $\lambda$ the universal functional is well
approximated by
\begin{eqnarray}
\nonumber
F[n_\lambda] &\approx & \lambda^{5/3}J[n]+\\
\nonumber
&&+ \lambda (T_s^{LDA}[n]+E_x^{LDA}[n]+E_c^{LDA}[n])+\\
\nonumber
&&+ \lambda^{1/3}(T_s^{GE2}[n]+E_x^{GE2}[n]+E_c^{GE2})+\\
\label{e42}
&&+ \lambda^{-1/3}T_s^{GE4}[n]\ ,
\end{eqnarray}
where the full LDA correlation energy is considered in Eq. (\ref{e42}).
Of course, in the limit $\lambda\rightarrow\infty$ the LDA
approximation of $F$ becomes exact (all the density parameters
vanish in this limit, by construction). In particular,
we have
\begin{equation}\label{e43}
\lim_{\lambda\rightarrow\infty}E_{xc}[n_\lambda]=E_{xc}^{LDA}[n_\infty],
\end{equation}
where $n_\infty(\R)=n(0)=const$.
Note that while this constraint is satisfied by most  non-empirical 
XC semilocal functionals (that recover the LDA for a constant density), 
it can be out of reach for some wavefunction methods, as the second-order 
perturbation theory of M\o ller-Plesset (MP2) and its modifications 
\cite{Grimme2,MP2}, or the random phase approximation (RPA) \cite{LP1,RPA1}. 
However, the sophisticated orbital-based inhomogeneous 
Singwi-Tosi-Land-Sj\"olander (ISTLS) method \cite{Dobson1,Dobson3}, 
as well the XC kernel of linear response time-dependent DFT
(in the context of the adiabatic-connection fluctuation-dissipation
theorem \cite{HG,GL,LP1}) of Ref. \cite{CP1}, are accurate for the 
uniform-electron-gas scaling. 

To provide an example of the utility of the uniform-electron-gas scaling
let us consider neutral jellium clusters with $N$ electrons and radius
$R=r_sN^{1/3}$, having the external potential
\begin{equation}\label{e44}
V^{jel}_{ext}(\R)=\left\{ \begin{array}{lll}
N(-\frac{3}{2R}+\frac{r^2}{2R^3}),     & r<R\\
-N \frac{1}{r},     & r\geq R,\\
\end{array}
\right.
\end{equation}
due to a positive background density
\begin{equation}\label{e45}
n_+(\R)=\left\{\begin{array}{ll}
3/4\pi r_s^3 & \;\;\;\;\; r<R\\
0 & \;\;\;\;\; r\geq R
\end{array}\right. \ .
\end{equation}
This external potential has no singularities, so the reduced gradients and
Laplacian of the density, are finite everywhere inside the bulk. 
Moreover, the values of the density parameters decrease for increasing
number of electrons $N$. Indeed, even for intermediate values of $N$, 
the density is slowly varying over a Fermi wavelength, so the 
extensions of Thomas-Fermi theory become accurate \cite{EP,APF,TPAFK,Eka}. 

Jellium clusters with different numbers of electrons may be thought 
therefore to be well described by the uniform-electron-gas scaling, 
since under the scaling procedure the number of electrons is 
changed to $\lambda N$, the local Seitz parameter is kept fixed 
to $r_s$, and the reduced gradients are decreased as $\lambda^{-1/3}$,
in full analogy to what happens in the jellium clusters. 
The relation between the uniform-electron-gas scaling and the 
jellium clusters can be in fact clearly recognized by a detailed
analysis of the cluster's electron densities.
\begin{figure}
\includegraphics[width=\columnwidth]{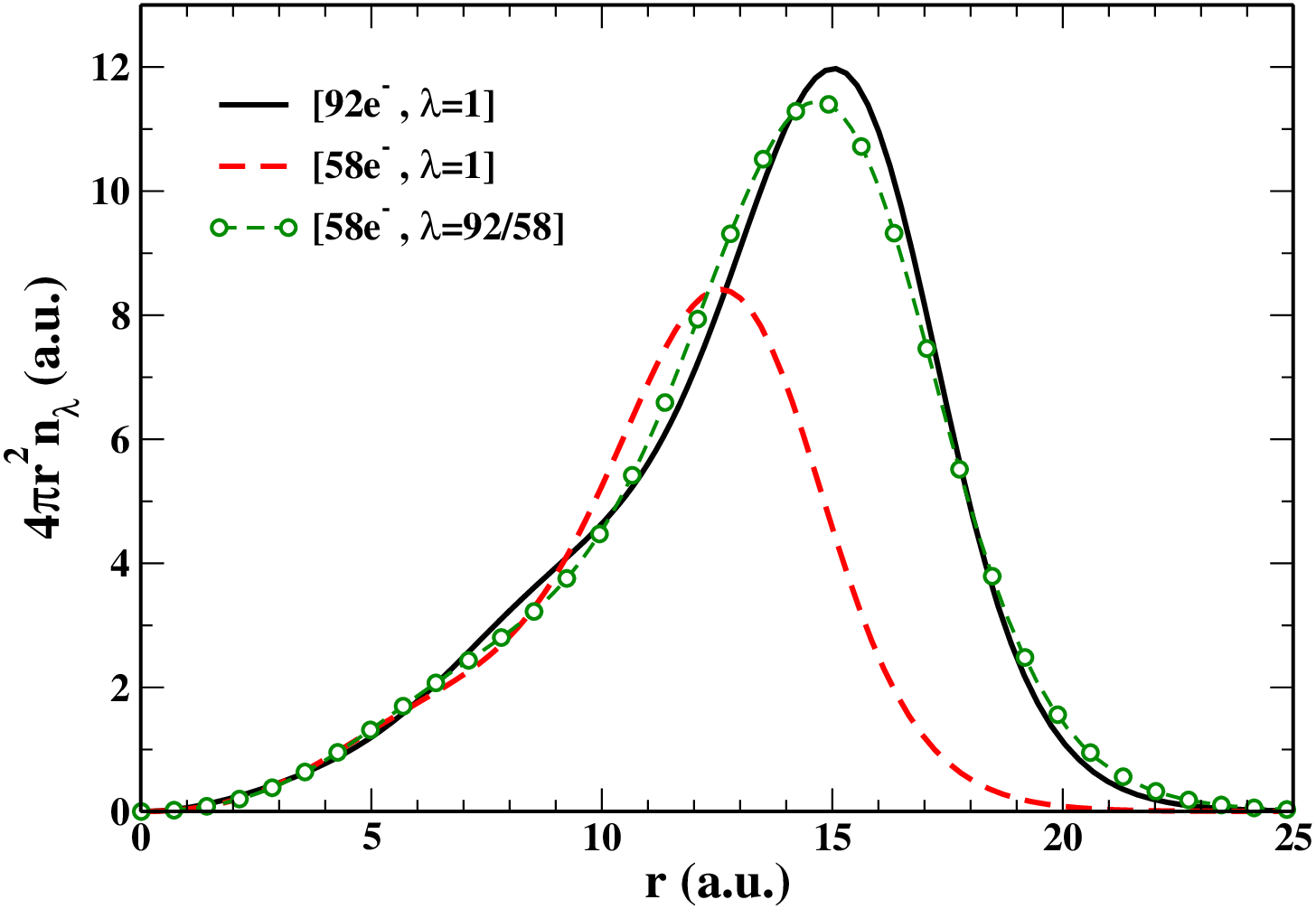}
\caption{(Color online) $4\pi r^2 n_\lambda$ versus the radial distance $r$ 
for the $58 e^-$ Na jellium cluster ($\lambda=1$ and $\lambda=92/58$), and for the $92 e^-$ 
Na jellium cluster ($\lambda=1$).The areas under the curves are the total number
of electrons (58 and 92 respectively).}
\label{f3}
\end{figure}
To this end, in Fig. \ref{f3} we show the densities of 
$58 e^-$ and $92 e^-$ Na jellium clusters, together with the 
UEG-scaled density of the $58 e^-$ cluster where a
value $\lambda=92/58$ was used. Remarkably, the scaled 
density agrees very well with the $92 e^-$ cluster density showing that
the variations of the physical properties of the different clusters with
$N$ can be well captured by the uniform-electron-gas scaling.
Note that the small differences between the scaled and the true density
are due to quantum oscillations \cite{Eka} that are not accounted by the
scaling procedure, but are of course included in the 
self-consistent Kohn-Sham scheme. These are however not
very relevant for the analysis that we consider henceforth.

The total energy of a jellium cluster can be written
\begin{equation}\label{e46}
E[n] = F[n] + E_{ext}[n] = F[n] + \int V^{jel}_{ext}(\R)n(\R)d\R\ .
\end{equation}
Therefore, using the uniform-electron-gas scaling for 
$\lambda\rightarrow\infty$ we can write
\begin{eqnarray}
\nonumber
E[n] & \approx & \lambda (T_s^{LDA}[n]+E_x^{LDA}[n]+E_c^{LDA}[n])+\\
\nonumber
&&+ \lambda^{1/3}(T_s^{GE2}[n]+E_x^{GE2}[n]+E_c^{GE2})+\\
\label{e47}
&&+ O(\lambda^{-1/3})\ ,
\end{eqnarray}
where we used the fact the Coulomb and external potentials cancel
each other in the limit of a large number of electrons.
This expression agrees well with the asymptotic expansion of 
the total energy of the clusters derived from the liquid drop model 
\cite{EP,APF,TPAFK} (recall that $N\sim\lambda$)
\begin{equation}\label{e48}  
E=\alpha\frac{4\pi r_s^3}{3}N+\sigma4\pi r_s^2 N^{2/3}+\gamma 2\pi r_s N^{1/3}, 
\end{equation}
where $\alpha$ is the volume (bulk) energy and $\sigma$ and $\gamma$ 
are the surface and curvature energies.
A comparison of the first terms (those scaling as $\lambda$) in 
Eqs. (\ref{e47}) and (\ref{e48}) shows in fact that the 
uniform-electron-gas scaling correctly yields 
$\alpha\sim T_s^{LDA}+E_x^{LDA}+E_c^{LDA}$.
Moreover, similar with the
previous Thomas-Fermi scaling (see Eq. (\ref{eee})), 
the second term in Eq. (\ref{e48}) is a quantum
oscillation term, that apparently can not be described by the simple UEG scaling, without
a careful analysis of the Friedel oscillations near the surface of the cluster.
Finally, the third term is related to the curvature energy 
\begin{equation}\label{e49} 
\gamma \sim T_s^{GE2}+E_x^{GE2}+E_c^{GE2}.
\end{equation}
This is an important result, because the real edge gas contains 
curvature corrections \cite{KM1}, that until now had not been addressed.


\subsection{Homogeneous density scaling ($\beta=0$)}
The homogeneous density scaling 
\cite{chan99,nagy05,cohenN,liu97,parr97,parr97_2,nagy06,morrison07} is obtained from
Eq. (\ref{e4}) by setting the parameter $\beta=0$.
Under this condition the scaling relations for the density
and the various density parameters are
\begin{eqnarray}
\label{e50}
n_\lambda(\R) = \lambda n(\R) \ & , & \ r_{s\lambda}(\R) = \lambda^{-1/3}r_s(\R)\\
\label{e51}
s_\lambda(\R) = \lambda^{-\frac{1}{3}}s\left(\R\right) \ & , 
& \ q_\lambda(\R) = \lambda^{-\frac{2}{3}}q\left(\R\right)\ ,\\ 
\label{e52}
t_\lambda(\R)  = \lambda^{-\frac{1}{6}}t\left(\R\right)\ & , 
& \ v_\lambda(\R) = \lambda^{-1/9}v\left(\R\right)\ .
\end{eqnarray}
For $\lambda\rightarrow\infty$ the high-density slowly-varying
limit is obtained and the universal functional is well 
approximated as
\begin{eqnarray}
\nonumber
F[n_\lambda] &\approx & \lambda^{2}J[n]+ \lambda^{5/3}T_s^{LDA}[n] + \\
\nonumber
&&+ \lambda^{4/3} E_x^{LDA}[n] + \lambda T_s^{GE2}[n] + \\
\label{e53}
&&+ \lambda^{2/3}E_x^{GE2}[n] + \lambda^{1/3}T_s^{GE4}[n]\ .
\end{eqnarray}
In this case the functional of Eq. (\ref{e53}) is dominated by the
classical Coulomb term, and the second leading term is the Thomas-Fermi 
kinetic energy. (Note that the nuclear energy 
grows 
only linearly with $\lambda$.) 

More importantly, the homogeneous scaling is a valuable tool to investigate
DFT, when the opposite limit, i.e. with $\lambda<1$, is considered.
For this case in fact several studies exist on the scaling 
properties and exact constraints of the kinetic and exchange 
energy functionals \cite{chan99,nagy05} as well as on the static
correlation treatment in DFT \cite{cohen08,cohen_science}.
In this work we focus instead on the role of the homogeneous scaling in
the determination of the delocalization error \cite{cohen_science} 
of the exchange(-correlation) functionals.
To this end we consider the simple H$^+_2$ dissociation problem, that is associated 
with the hydrogen atom with fractional charge \cite{cohen08}, and
compute
\begin{equation}\label{e54}
\Delta E_H(q) = E_H - E_{H^{q}} - E_{H^{(1-q)}}\ ,
\end{equation}
where $E_H=E_{xc}+J$ for the hydrogen atom and $0\leq q\leq 1$ is
the partial electronic charge. The quantity $\Delta E_H$ represents thus the
Coulomb and XC energy difference between the dissociation of H$^+_2$ into one
hydrogen atom plus one proton and that of the dissociation into
two hydrogen atoms with fractional electron charge $q$ and $1-q$.
For the exact exchange-(correlation) functional it shall be zero
at any value of $q$. However, due to the one-electron self-interaction error
\cite{zhang98}, for approximated XC functionals $\Delta E_H>0$ 
for any $0<q<1$, indicating that a fractional dissociation is 
favorable with respect to the exact one.

\begin{figure}
\includegraphics[width=\columnwidth]{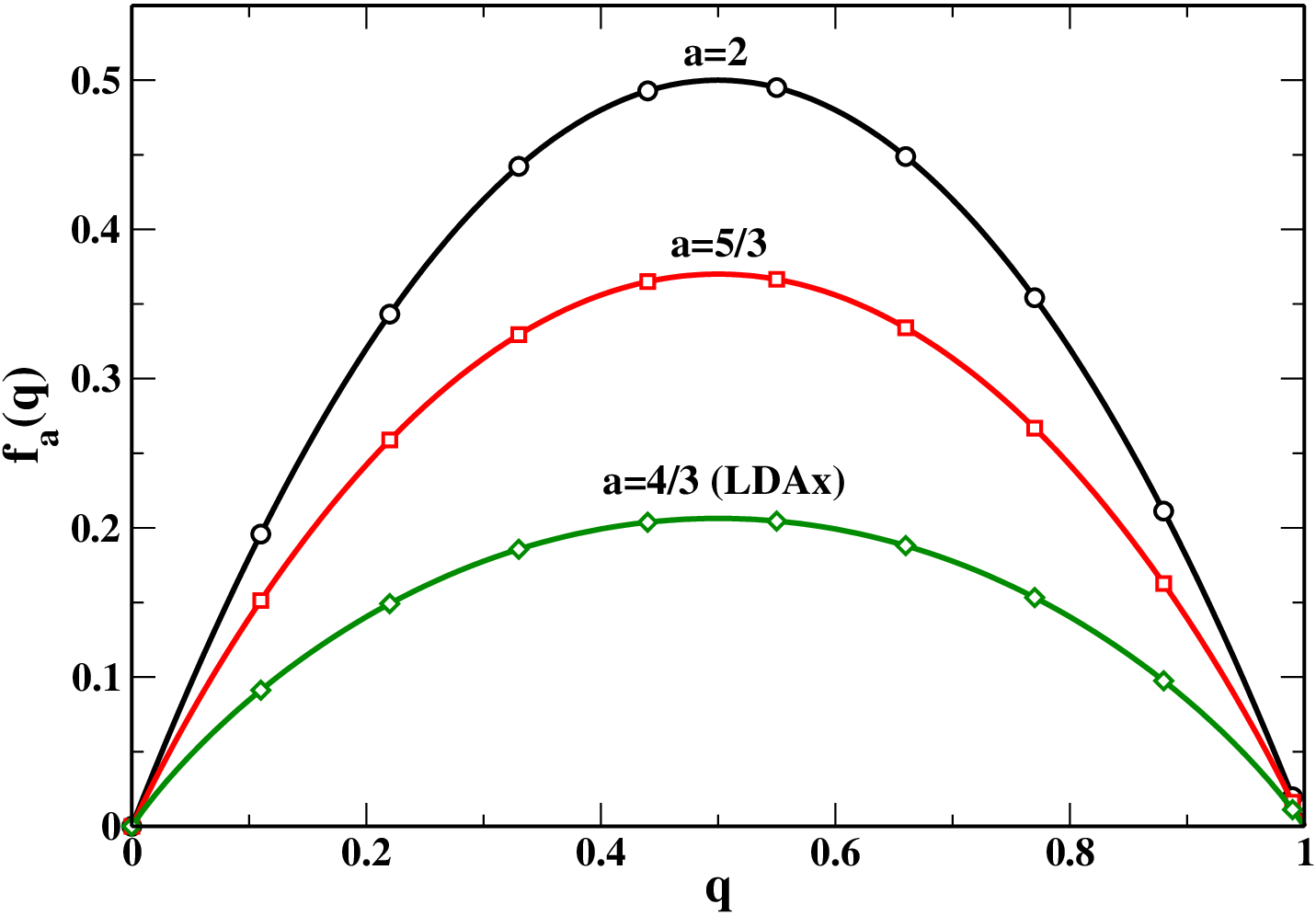}
\caption{\label{ffactor}(Color online) Plot of $f_a(q)=1-q^a-(1-q)^a$ for several values of the parameter $a$.}
\end{figure}

\begin{table}[b]
\begin{center}
\caption{\label{tab_sie} Self-interaction error (SIE) as defined in 
Eq. (\ref{e56}) for different popular exchange and exchange-correlation functionals. 
All values are computed for the hydrogen density and expressed in mHartree.}
\begin{ruledtabular}
\begin{tabular}{llrcllr}
\multicolumn{3}{c}{X-only functionals} & & \multicolumn{3}{c}{XC functionals} \\
\cline{1-3}\cline{5-7}
Functional & Ref. & SIE & $\qquad$ &Functional & Ref. & SIE \\
LDAx   & \cite{lda1,lda2} & 44.5 & & LDA  &\cite{lda1,lda2,pw92} & 22.3 \\ 
PBEx   & \cite{PBE}       &  6.6 & & PBE  & \cite{PBE}           &  5.8 \\
APBEx  & \cite{APBE}      &  1.8 & & APBE & \cite{APBE}          & -3.4 \\
revPBEx& \cite{revpbe}    &  2.0 & & revPBE& \cite{revpbe}       & -4.0 \\
PBEsolx& \cite{pbesol}    & 19.8 & & PBEsol& \cite{pbesol}       & 11.9 \\
       &                  &      & & zPBEsol&\cite{pbesol,zeta}  &  0.0 \\
PBEintx& \cite{pbeint}    & 16.9 & & PBEint & \cite{pbeint}      &  9.7 \\
       &                  &      & & zPBEint&\cite{pbeint,zeta}  &  0.1 \\
B88    & \cite{b88}       &  2.7 & &        &                    &  \\
\end{tabular}
\end{ruledtabular}
\end{center}
\end{table}
For a generic exchange functional we can assume (see Section
\ref{sec5}) the scaling $E_x^{any}[n_\lambda]=\lambda^aE_x^{any}[n]$,
with $a\lesssim 2$. Thus, Eq. (\ref{e54}) becomes
\begin{eqnarray}
\nonumber
\Delta E_H(q) & = & \left(1-q^2-(1-q)^2\right)J[n_H] + \\
\label{e55}
&& + \left(1-q^a-(1-q)^a\right)E_x^{any}[n_H]\ ,
\end{eqnarray}
with $n_H$ the density of the hydrogen atom and $J[n_H]=0.3125$ Ha. 
Note that in this case
correlation plays no role, since for $q\leq 1$ the correlation
energy is zero. Nevertheless for semilocal functionals,
which are not self-correlation free, it is also possible to consider 
an $E_{xc}^{any}$ in place of $E_x^{any}$.
Now, in analogy with Ref. \onlinecite{zhang98}, we can define the 
self-interaction error as
\begin{equation}\label{e56}
SIE[n] = J[n] + E_{xc}[n]\ .
\end{equation}
Equation (\ref{e55}) then becomes
\begin{equation}\label{e57}
\Delta E_H(q) = \left(f_2(q)-f_a(q)\right)J[n_H] + f_a(q)SIE[n_H]\ ,
\end{equation}
with $f_a(q)=1-q^a-(1-q)^a$. A plot of $f_a$ for several values of
$a$ is provided in Fig. \ref{ffactor}.

Equation (\ref{e57}) shows that $\Delta E_H$ is always positive whenever
the self-interaction error is not zero and the scaling 
behavior of the functional differs from the exact one 
($E_x[n_\lambda]=\lambda^2E_x[n]$), since $J\gg SIE$ and
$SIE$ is in general positive.
Moreover, due to the form of the function $f_a$, a symmetric fractional
dissociation is always favored. The most important result of Eq. (\ref{e57})
is however the fact that the delocalization error (or dissociation 
error in H$^+_2$) has a twofold origin and only the second term on 
the right hand side depends on the self-interaction error for the 
hydrogen atom. The remaining contribution traces back only to the wrong
behavior of the functional under the homogeneous scaling. Note also that
the first term might be the dominant one since the SIE is typically
one order of magnitude smaller than the Coulomb energy in the H atom
(see Table \ref{tab_sie} for the value of $SIE[n_H]$ for some popular 
functionals).

This result shows the importance of the homogeneous density scaling
in the development of accurate XC functionals. In fact, several functionals
were constructed to minimize the self-interaction error of the
hydrogen atom \cite{delcampo12}, but no semilocal approximation exists that
provides the correct scaling behavior under uniform density scaling.


\subsection{Fractional scaling ($\beta=1$)}
Let consider the disintegration of
the hydrogen atom into two neutral atoms
having a fractional number of electrons and corresponding fractional
nuclear charge.
The electron density of the hydrogen atom with a fractional number of
electrons $q$, must be considered as the ensemble density given by the
superposition of the density of an hydrogen atom with one electron and
nuclear charge $q$, with weight $q$, and that of one hydrogen atom with no
electrons and nuclear charge $q$, with weight zero (this latter thus does
not contribute).
Recalling that for the hydrogen atom, a scaling of the nuclear charge corresponds to an
uniform scaling of the density, we find that
\begin{equation}\label{e59}
n_q(\R) = q^4n_1(q\R)\ .
\end{equation}

Generalizing the result of Eq. (\ref{e59}), we define the fractional scaling
as the scaling obtained from the family of relations given in Eq. (\ref{e4})
when $\beta=1$. Under this scaling the density and the various density
parameters behave as
\begin{eqnarray}
\label{e60}
n_\lambda(\R) = \lambda^4 n(\lambda\R) 
\ & , & \ r_{s\lambda}(\R) = \lambda^{-4/3}r_s(\lambda\R)\\
\label{e61}
s_\lambda(\R) = \lambda^{-\frac{1}{3}}s\left(\lambda\R\right) 
\ & , & \ q_\lambda(\R) = \lambda^{-\frac{2}{3}}q\left(\lambda\R\right)\ ,\\ 
\label{e62}
t_\lambda(\R)  = \lambda^{\frac{1}{3}}t\left(\lambda\R\right)\ 
& , & \ v_\lambda(\R) = \lambda^{5/9}v\left(\lambda\R\right)\ .
\end{eqnarray}
In the limit $\lambda\rightarrow\infty$ the system is scaled towards
the high-density limit ($r_s\rightarrow 0$) and the universal functional
is well approximated as
\begin{eqnarray}
\nonumber
F[n_\lambda] &\approx & \lambda^{11/3}T_s^{LDA}[n] + \lambda^{3}\left(J[n]+T_s^{GE2}[n]\right)+  \\
\nonumber
&&+ \lambda^{7/3} \left(E_x^{LDA}[n]+ T_s^{GE4}[n]\right) + \\
\label{e63}
&&+ \lambda^{5/3}E_x^{GE2}[n] \ .
\end{eqnarray}
Note however that in this limit the reduced gradients for the correlation
are not small (and thus the correlation energy is not important).
 Eq. (\ref{e63}) shows a particular 
classical behavior 
of the electrons: their kinetic energies become dominant over the 
classical Coulomb and exchange energies. Thus, in the limit 
$\lambda\rightarrow\infty$, the electronic system shows similarities with 
a gas of non-interacting particles, with non-uniform density 
$n_\lambda(\R)=\lambda^4n(\lambda\R)$.  

\begin{figure}
\includegraphics[width=\columnwidth]{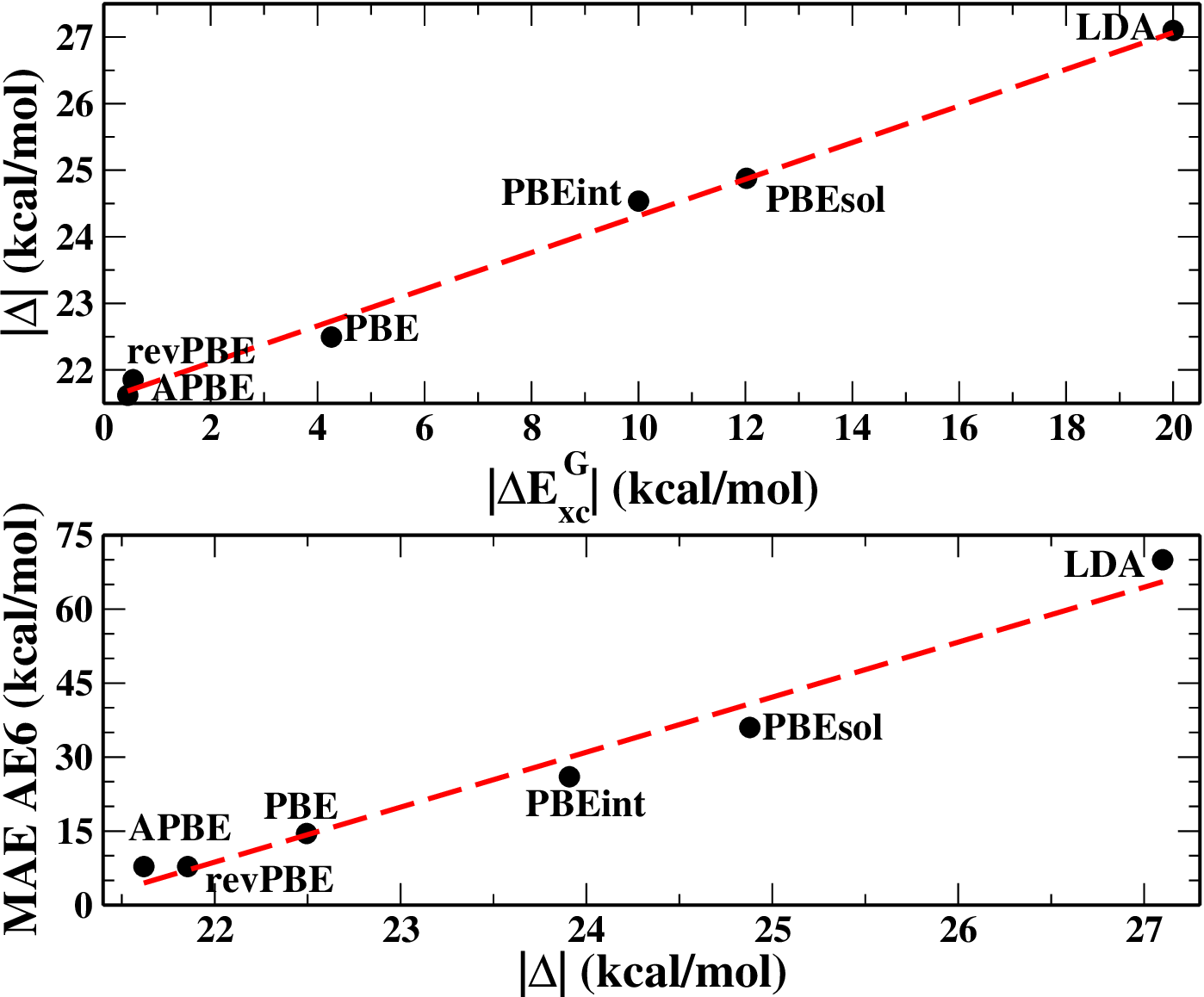}
\caption{\label{fig4}(Color online) Gaussian density XC error versus disintegration error
$\Delta$(upper panel) and disintegration error $\Delta$ versus
mean absolute errors of the AE6 test (lower panel), for
several functionals (LDA \cite{KS}, PBE \cite{PBE}, revPBE \cite{revpbe},
APBE \cite{APBE}, PBEint \cite{pbeint}, and PBEsol \cite{pbesol}.
The dashed lines are linear interpolations of the data.}
\end{figure}

In the opposite limit ($\lambda<1$), the fractional scaling describes, as
discussed above, the scaling towards a fractional atom. In fact, in this case
the system is correctly scaled towards the low-density limit with the
reduced gradient and Laplacian for the kinetic and exchange energy becoming 
large.
To strengthen the significance of such a scaling we consider its application
to the disintegration of an atom, as introduced above.
For simplicity we consider the disintegration of a hydrogen 
atom into two neutral atoms having a fractional number of electrons 
$q$ and $1-q$, respectively, and corresponding fractional nuclear charge.
We define the XC disintegration energy as
\begin{equation}\label{e64}
M(q)=E_{xc}(1)-E_{xc}(q)-E_{xc}(1-q),
\end{equation}
where $E_{xc}(q)$ denotes the XC energy of the H atom with fractional
electron number and nuclear charge $q$. Note that correlation will only
play a role when approximate non-self-interaction-free DFT functionals
are considered ($M^{DFT}$), while only exchange will contribute in the 
computation of $M^{exact}$. 
This quantity is very important since the accuracy of any
GGA functional in computing the values of $M(q)$ is directly
related to its ability to predict good atomization energies, 
because both processes (disintegration and atomization) preserve 
the total number of electrons. 
This fact is clearly shown in Fig. \ref{fig4} where it can be noted
the linear relation between the mean absolute error (MAE) on
the computation of atomization energies of organic molecules
(we considered here the AE6 test \cite{AE6}, which is representative
for organic molecule atomization energies) and the 
disintegration error $\Delta$ defined as
\begin{equation}\label{e65}
\Delta = \int_0^1\left[M^{DFT}(q)-M^{exact}(q)\right]dq\ ,
\end{equation}
for several representative XC functionals.  
Of course, a linear relation is obtained as well with the errors
on the XC energy of a Gaussian one-electron density, which was shown to
be a model system for atomization energies \cite{zeta2,zeta}.

The disintegration error can be written alternatively
\begin{equation}\label{e66}
\Delta = E_{xc}^{DFT}(1)-E_{xc}^{exact}(1) -2\int_0^1\left[E_{xc}^{DFT}(q)-E_{xc}^{exact}(q)\right]dq\ .
\end{equation}
For the exact energy we have $E_{xc}^{exact}(q)=E^{exact}_x(q)=q^3E_x^{exact}(1)$,
while for the approximate functionals we found numerically
(see section \ref{sec5}) $E_{xc}^{DFT}(q) = q^aE_{XC}^{DFT}(1)$ with
$a\sim$2.3. Substituting into Eq. (\ref{e66}) we obtain
\begin{equation}\label{e67}
\Delta \sim 0.39 E_{xc}^{DFT}(1) - \frac{1}{2}E_{xc}^{exact}(1)\ .
\end{equation}
Equation (\ref{e67}) shows that: (i) the hydrogen disintegration
problem is not related to the hydrogen one-electron density,
as might appear at a first sight. Indeed, for a GGA which is exact for the
hydrogen atom we would have $\Delta\sim$ 22 kcal/mol.
In fact, Fig. \ref{fig4} shows that the hydrogen atom
disintegration is rather linearly related to the Gaussian one-electron
density.
(ii) The error $\Delta$ is determined basically from the behavior
of the functional for $q$=1.
(iii) To minimize the disintegration error (and thus yield good
atomization energies), $E_{xc}^{DFT}$ must be
about 10-20\% larger than $E_{xc}^{exact}$ for the hydrogen atom;
this condition is well met by GGA functionals accurate for
atomization energies (APBE \cite{APBE}, revPBE \cite{revpbe}), and is also well satisfied 
by the recently proposed zvPBEsol, and zvPBEint \cite{zeta2}. 
However, this result shows that in fact all these functionals base their
performace on an error cancellation effect which is made inevitable by their
inability to respect the fractional scaling.

Thus, the fractional scaling appears as an important exact constraint that
XC functionals should respect. At present all popular approximations to the XC
energy are only designed to respect the uniform scaling and are unable to
fulfill the fractional scaling relations.


\subsection{Strong-correlation scaling ($\beta=-1$)}
For $\beta<-1/3$, and in the semiclassical limit $\lambda\rightarrow\infty$, 
$E_x^{LDA}[n_\lambda]=\lambda^{\beta+4/3}E_x^{LDA}[n]$ starts to 
dominate over the Thomas-Fermi kinetic energy 
$T_s^{LDA}[n_\lambda]=\lambda^{2\beta+5/3}T_s^{LDA}[n]$, and thus 
the system may reach a strongly-correlated limit.
Let us briefly consider the case $\beta=-1$.

With this choice the scaling of the density and the reduced gradients are
\begin{eqnarray}
\label{eq35}
n_\lambda(\R) = \lambda^{-2}n(\lambda^{-1}\R) \ & , & \ r_{s\lambda}(\R) = \lambda^{2/3}r_s(\lambda^{-1}\R),\\
\label{eq36}
s_\lambda(\R) = \lambda^{-1/3}s(\lambda^{-1}\R) \ & , & \ q_\lambda(\R) = \lambda^{-2/3}q(\lambda^{-1}\R)\ ,\\
\label{eq37}
t_\lambda(\R)  = \lambda^{-2/3}t(\lambda^{-1}\R)\ & , & \ v_\lambda(\R) = \lambda^{-7/9}v(\lambda^{-1}\R)\ .
\end{eqnarray}
and thus, under this scaling with $\lambda\rightarrow\infty$, the system 
is in a slowly-varying, low-density limit. Thus, 
our approximation for the universal functional becomes
\begin{eqnarray}
\nonumber
F[n_\lambda] & \approx & \lambda J[n]+\lambda^{1/3}\left(E_x^{LDA}[n]+E_c^{LDA}[n]\right)\\
\nonumber
&&\lambda^{-1/3}\left(T_s^{LDA}[n]+E_x^{GE2}[n]\right)+\\
&&+\lambda^{-1}T_s^{GE2}[n]+\lambda^{-5/3}T_s^{GE4}[n]\ ,
\label{eq34}
\end{eqnarray}
which shows that the kinetic energy of the electrons, is much smaller 
than the Coulomb interaction, and the system resembles the features 
of a Wigner crystal, being in a strongly-correlated limit. 
Note that in this limit, the LDA correlation energy scales as 
$E_c^{LDA}[n_\lambda]=\lambda^{1/3}E_c^{LDA}[n]$, being 
as important as the exchange part. We recall that such a Wigner crystal 
is well described by a semilocal functional (see Eq. (23) of Ref. \cite{PK1}) derived 
from the point-charge-and-continuum (PC) model \cite{PK1}. 
Note that the PC model was incorporated in 
high-level 
methods (e.g. ISI method of Ref. \cite{ISI}), and recent excellent work has 
been done for further development of the PC model, see Refs. 
\cite{Vignale1,Kieron1},
as well for a DFT of strongly-correlated systems \cite{Vignale2,Paola1}. 
Thus, further study of this scaling can be important. 


%
\begin{table}
\begin{center}
\caption{\label{tab_seff} Effective scaling order $s_{eff}$
(see Eq. (\ref{e68}) for several
semilocal functionals at $\beta=0$ and $\beta=1$.
The last line reports the reference value for the exact exchange
functional. The second column reports the reference to the appropriate 
literature for each functional.}
\begin{ruledtabular}
\begin{tabular}{llrr}
Functional & Ref.  & $\beta=0$ &  $\beta=1$ \\
\hline
\multicolumn{4}{c}{Exchange-only functionals} \\
LDAx   & \cite{lda1,lda2} &  1.333 &  2.666 \\ 
B88    & \cite{b88}       &  1.254 &  2.206 \\
PBEx   & \cite{PBE}       &  1.266 &  2.323 \\
APBEx  & \cite{APBE}      &  1.262 &  2.316 \\
revPBEx& \cite{revpbe}    &  1.252 &  2.252 \\
PBEsolx& \cite{pbesol}    &  1.280 &  2.356 \\
PBEintx& \cite{pbeint}    &  1.272 &  2.329 \\
TPSSx  & \cite{tpss}      &  1.297 &  2.399 \\
revTPSSx & \cite{revtpss} &  1.309 &  2.439 \\
Exact  &                  &  2.000 &  3.000 \\
       &                  &       &             \\
\multicolumn{4}{c}{Exchange-correlation functionals} \\
LDA&\cite{lda1,lda2,pw92} &  1.317 &  2.517 \\
PBE    & \cite{PBE}       &  1.272 &  2.342 \\
APBE   & \cite{APBE}      &  1.268 &  2.333 \\
revPBE & \cite{revpbe}    &  1.258 &  2.271 \\
PBEsol & \cite{pbesol}    &  1.287 &  2.378 \\
zPBEsol&\cite{pbesol,zeta}&  1.268 &  2.316 \\
PBEint & \cite{pbeint}    &  1.279 &  2.351 \\
zPBEint&\cite{pbeint,zeta}&  1.261 &  2.289 \\  
Exact  &                  &  2.000 &  3.000 \\
\end{tabular}
\end{ruledtabular}
\end{center}
\end{table}
%


\section{Effective scaling for semilocal functionals}
\label{sec5}
In this section we consider an assessment of semilocal XC 
density functionals for various scalings discussed above.
In previous sections we discussed the scaling properties
of several exact energy functionals under the
scaling transformations of the type of Eq. (\ref{e4}) and 
we proved the utility of such scaling relations in various contexts.
At the same time, we noted that these scaling relations
are not respected by approximate functionals
even at the LDA or second-order gradient-corrected
level (the von Weizs\"{a}cker kinetic energy functional
being one exception). 
The situation is even worst for generalized gradient approximations
which indeed do not have a well defined scaling behavior under the
scaling transformations of Eq. (\ref{e4}). 

Nevertheless, it can be seen that most XC functionals in fact
display an effective scaling $E_{xc}[n_\lambda] = \lambda^aE_{xc}[n]$.
Solving this equation for the parameter $a$, we find 
$a=[\ln(|E_{xc}[n_\lambda]|)-\ln(|E_{xc}[n]|)]/\ln(\lambda)$.
Therefore, it is conceivable to define an effective scaling order
for a generic functional $E_{xc}$ as 
\begin{equation}\label{e68}
s_{eff} = \int_0^1\frac{\ln\left(\left|E_{xc}[n_H]\right|\right)-\ln\left(\left|
E_{xc}[n_{H\lambda}]\right|\right)}{\ln(\lambda)}d\lambda\ .
\end{equation}
Because our interest for the scaling of general exchange(-correlation) 
functionals
in this work was motivated by the homogeneous and fractional scaling,
we restricted our definition to use of hydrogen density $n_H=\exp(-2r)/\pi$
and the interval $\lambda\in(0:1)$, where we can compare with exact results. 

The effective scaling order provides a
measure for the scaling behavior of different functionals, 
resembling in this respect the effective homogeneity of XC functionals
\cite{zhao94,tozer98,borgoo12}. Of course, for functionals 
having a well defined scaling behavior (e.g. LDAx), the
effective scaling order will coincide with the analytic scaling exponent.
For other functionals it will provide a measure of the effective
scaling behavior, so that the deviations of the effective scaling order 
from the true value could give an estimation of the accuracy of the functional
to fulfill the scaling relation.
We note that for the latter cases the integrand of Eq. (\ref{e68}) was
always found to be almost constant over the entire integration interval
(except very close to the boundaries; note however that the function
is integrable over the given range), showing the robustness of our 
definition. 

In Table \ref{tab_seff} we report the values of $s_{eff}$ for
several exchange and exchange-correlation functionals at
$\beta=0$ and $\beta=1$.
An inspection of the data shows that all the functionals perform
similarly and quite differently from the exact reference,
that in this case is the exact Kohn-Sham exchange, since for 
$n=n_H$ and $\lambda<1$ there is no correlation.
Remarkably, the best scaling behavior is found for LDA exchange,
while slightly worst results are obtained for GGA functionals. 
The use of meta-GGA functionals, as
TPSS \cite{tpss} or revTPSS \cite{revtpss}, which
are constructed taking into account the physics
of one-electron systems, is found finally to bring a slight improvement
in the effective scaling behavior.
Moreover, at the GGA level
the addition of approximate correlation seems 
to bring some small improvement in the scaling behavior,
in line with the fact that indeed semilocal DFT functionals
are not really exchange  or correlation functionals but rather 
rely on an heavy error compensation between the two.
We recall instead that meta-GGA functionals are 
one-electron-self-correlation free.


\section{Conclusions}
\label{sec6}
In summary, we have investigated the scaling with 
variable particle number of the form of Eq. (\ref{e4}): 
$n_\lambda(\R)=\lambda^{3\beta+1}\; n(\lambda^\beta \R)$.
For such scaling transformations we provided a formal
definition within the ensemble formalism of DFT and  studied the
basic features, also in relation to the scaling properties of different
important density functionals.

The density scalings defined in Eq. (\ref{e4}),
spans an impressive set of physical properties: in the limit of
large $\lambda$ they are crucial
for semiclassical theory of many-electron systems (e.g.
Thomas-Fermi scalings is related to atoms, Uniform-electron-gas
scaling is related to metallic clusters, strong-correlation scaling
is related to
Wigner crystals), whereas in the limit of small $\lambda$ they are
connected with the physics of small systems with fractional particle number,
and to self-interaction errors.

The here proposed uniform-electron-gas scaling ($\beta=-1/3$) is 
the right basic concept for jellium clusters. Simple scaling 
manipulations showed that the curvature corrections are in fact 
related to the second-order gradient expansion. On the other hand, the
surface corrections (described by the Airy gas model \cite{KM1}), 
are quantum oscillations terms. By analogy with the recent work 
on the semiclassical atom \cite{ELCB,LCPB,EB09}, a modified second-order 
gradient expansion (MGE2) can be constructed for jellium clusters, 
in order to recover the exact surface corrections.  
However, we expect that such a MGE2 will be very close to the regular GE2
that is accurate for surfaces of simple metals 
when the Kohn-Sham densities are used \cite{pbesol}. 

Moreover, the idea of MGE2, that can account for the principal quantum corrections,
 can be generalized for any $\beta$ (in 
the limit of large $\lambda$). Such a $\beta$MGE2 will be very useful
especially for the strongly-correlated scaling ($\beta=-1$) where 
the LDA term is not exact (in the limit $\lambda\rightarrow\infty$) due 
to the self-interaction problem in Wigner crystals \cite{PK1}. 
For example, $\beta$MGE2 for the exchange energy may have the form
\begin{equation}\label{efinal}
E_x^{\beta \rm{MGE2}}[n]=\int d\R \; n\; \epsilon_x^{LDA}f(\beta)(1+\mu(\beta)s^2),
\end{equation}
where $f(\beta)$ and $\mu(\beta)$ should be derived in further 
investigations. (For the Thomas-Fermi scaling $f(\beta=1/3)=1$ and $\mu(\beta=1/3)=0.26$ \cite{EB09};
whereas the regular GE2 has $f=1$ and $\mu=0.12346$).

Finally we have shown the usefulness of the here proposed fractional scaling  
($\beta=1$) for the atomization energies of molecules. Recently, it has been 
derived an atomization energy constraint (i.e. minimization of an entropy-like 
function for an ensemble one-electron density models) \cite{zeta,zeta2}. This 
constraint was derived from an empirical observation relating
errors in the model one-electron densities to errors in the atomization
energies, of popular GGAs (see Fig. 2 of Ref. \cite{zeta}); and from the physical explanation
 that one-electron densities are simple models for simple
bonding regions, where iso-orbital regime can be significant.
(see Fig. 1 of Ref. \cite{zeta}, and the corresponding discussion).
Using the fractional scaling
($\beta=1$), we have better explained the significance of 
one-electron Gaussian model for atomization energy of molecules and 
disintegration of the hydrogen atom (see Eq. (\ref{e67}) and its related 
discussion). 

We recall that the semilocal exchange hole models 
satisfy the sum rule for systems with integer number of electrons \cite{solhole}, 
but violate the exchange hole sum rule in case of fractional number 
of electrons, and thus predicting too-negative energies for such systems \cite{perdew12}. 
The here proposed effective scaling (see Section V), measures in fact 
the functional accuracy for systems with fractional particle number,
and can be used to develop (and test) new better approximations.

Our work provides a deeper insight into the relevance
of the scaling relations having the form defined in Eq. (\ref{e4}) and
highlights the importance of these scaling relations in DFT.
In particular, Eqs. (\ref{e22})-(\ref{eee22}) provide useful
scaling relations with
varying particle number that are important constraints in the construction
of approximate exchange-correlation (or noninteracting kinetic) functionals.
Moreover, for the $\lambda\rightarrow 0$ limit, Eqs. (\ref{ee22}) and
(\ref{ee24}) are exact constraints for the Kohn-Sham 
exchange and kinetic energy functionals, respectively and the 
effective scaling order (Eq. (\ref{e68})) is a more general
requirement, which is relevant for the SIE problem.
Unfortunately, no such explicit expressions exist instead for the 
$\lambda\rightarrow \infty$ limit, where however the recovery of the 
semiclassical atom physics was shown to be an important condition.

\begin{acknowledgment}
We thank F. Della Sala for useful discussions and acknowledge
funding by the European Research Council (ERC) 
Starting Grant FP7 Project DEDOM, grant agreement no. 207441.
\end{acknowledgment}

\appendix
\section{Scaling relations for Kohn-Sham kinetic and exchange energies for one electron systems}
\label{appa}
In this appendix we consider briefly the special case of a fractional
number of particles (with $N\leq 1$) and $\lambda\leq 1$. In this case, using the formalism 
of
Section \ref{sec2} we can derive,
in line with Ref. \onlinecite{PL1}, the general scaling properties for
the non-interacting kinetic energy and the Kohn-Sham exchange functionals.
Under coordinate 
scaling the density operator is transformed into another valid density 
operator and because of the simple scaling behavior of the
statistical weight we have
$\hat{\Gamma}_\lambda(\R) =\lambda\hat{\Gamma}(\lambda^\beta\R)$.
At this point we can write $\Tr{\hat{\Gamma}_\lambda\hat{T}}=
\lambda^{2\beta+1}\Tr{\hat{\Gamma}\hat{T}}$, where we used the fact that
$\hat{T}(\R/\lambda^\beta)=\lambda^{2\beta}\hat{T}(\R)$, and we
note that, for any $\lambda$, if $\Tr{\hat{\Gamma}\hat{T}}$ is a minimum
so must be $\Tr{\hat{\Gamma}_\lambda\hat{T}}$. Hence,
\begin{equation}\label{a22}
T_s[n_\lambda]=\lambda^{2\beta+1}T_s[n] \ .
\end{equation}
In a similar way it can be proved that
\begin{equation}\label{a24}
E_x[n_\lambda]  =  \lambda^{\beta+2}E_x[n] \ .
\end{equation}

\end{document}